\documentclass[12pt]{article}%
\usepackage{amssymb,amsbsy,url}

\usepackage{amsfonts,amsbsy}
\usepackage{amssymb}
\usepackage{graphicx}
\usepackage{amsfonts}%

\newcommand{\be}{\begin{equation}}
\newcommand{\ee}{\end{equation}}
\newcommand{\ba}{\begin{array}}
\newcommand{\ea}{\end{array}}

\newcommand{\fl}{}
\newcommand{\ds}{\displaystyle}
\newtheorem{theorem}{Theorem}

\begin{document}

\title{\protect\vspace*{-17mm}\bf Natural coordinates for a class of Benenti systems}

\author{Maciej B\l aszak$\dagger$ and Artur Sergyeyev$\ddagger$%
\\
$\dagger$\
Institute of Physics, A. Mickiewicz
University,\\ Umultowska 85, 61-614 Pozna\'{n}, Poland
\\
$\ddagger$\
Silesian University in Opava, Mathematical Institute,\\ Na
Rybn\'\i{}\v{c}ku 1, 746\,01 Opava, Czech Republic\\
E-mail: {\tt blaszakm@amu.edu.pl} and {\tt Artur.Sergyeyev@math.slu.cz}}
\date{}
\maketitle

\begin{abstract}
We present explicit formulas for the coordinates in which the
Hamiltonians of the Benenti systems with flat metrics take natural
form and the metrics in question are represented by constant
diagonal matrices.

\medskip

\hspace*{2.69cm} separation of variables.

{\bf PACS 2006}: 02.40.Ky, 02.30.Ik, 45.20.Jj
\end{abstract}

\section*{Introduction}
The separation of variables (SoV) undoubtedly is the most powerful
technique for integrating the equations describing classical
dynamics, namely the Hamilton--Jacobi equation. There are two basic
ways of using SoV. The first one consists in considering a class of
Liouville integrable Hamiltonian systems of physical or mathematical
significance and subsequent finding the separation coordinates in
order to integrate these systems by quadratures. This way is
extensively described in the existing literature on the subject, see
e.g.\ \cite{mb, Sklyanin, kal, mil} and references therein.
\looseness=-1

The second way, somewhat less explored, starts with the
systems written in the separation coordinates, so the corresponding
classical dynamics is \emph{a priori} integrable, and consists in finding new
(natural) coordinates in which the systems in question may admit a
physical interpretation. In the present paper we follow this
way and find natural coordinates for a subset of a very large class of separable
systems described by Benenti \cite{ben93}. We start with the
systems in question written down in the separation coordinates and present a sequence
of transformations leading to the natural coordinates,
i.e., orthogonal coordinates with a constant metric tensor and the corresponding momenta,
see Theorems~\ref{flath} and \ref{orth} below for details.

First of all, recall the basic aspects of modern geometric
approach to the separation of variables. Consider the following
class of separation (spectral) curves \cite{Sklyanin, mac2005}
\begin{equation}
H_{1}\lambda^{n-1}+H_{2}\lambda^{n-2}+\cdots+H_{n}=\frac{1}{2}\lambda^{m}%
\mu^{2}+\lambda^{k},\quad m,k\in\mathbb{Z},\quad n\in
\mathbb{N}.\label{SC1}%
\end{equation}
Relations (\ref{SC1}) contain complete information about a large
class of the so-called Benenti systems \cite{ben93, ben97}.
The separable systems from this class, labelled by the indices $m$ and $k$,
describe one-particle dynamics on Riemannian manifolds and belong to
a yet larger class of the
classical St\"{a}ckel systems. Taking $n$ copies of the curve
(\ref{SC1}) with variables $(\lambda,\mu)$ labelled within each copy
as $(\lambda^{i},\mu_{i})$ yields a system of $n$ separation
relations in the form of $n$ equations linear in the $H_{i}$.
Solving these equations for fixed $m$ and $k$ yields $n$ functions $H_{r}^{(m,k)}%
=H_{r}^{(m,k)}(\lambda,\mu)$ of the St\"{a}ckel form%
\begin{equation}
H_{r}^{(m,k)}=\frac{1}{2}\mu^{T}K_{r}G_{m}\mu+V_{r}^{(k)}\quad
r=1,\ldots,n,\quad m,k\in\mathbb{Z},\label{Ham}%
\end{equation}
where $\lambda=(\lambda^{1},\ldots,\lambda^{n})^{T}$ and
$\mu =(\mu_{1},\ldots,\mu_{n})^{T}$.
In turn, for fixed $m$ and $k$ the
functions (\ref{Ham}) can be interpreted as $n$ Hamiltonians on the phase space
$T^{\ast}\mathcal{Q}$, where
$\mathcal{Q}=\{(\lambda^1,\dots,\lambda^n)\in\mathbb{R}^{n}|\lambda^1 < \lambda^2<\cdots <\lambda^n\}$
is endowed with the contravariant metric tensor $G_{m}$ (even more broadly,
$H_{r}^{(m,k)}$ can be interpreted
as Hamiltonians on $T^{\ast}\mathcal{M}$, the cotangent
bundle for a pseudo-Riemannian manifold
$\mathcal{M}$ with the contravariant metric $G_m$).
These Hamiltonians are in involution with respect to the canonical
Poisson bracket on $\mathbb{R}^{2n}$.
Moreover, the Hamiltonians in question are
separable because they satisfy
the St\"{a}ckel separation relations (\ref{SC1})
by construction~\cite{mac2005}. The objects $K_{r}$ in (\ref{Ham})
can be interpreted as Killing tensors of type $(1,1)$ on $\mathcal{Q}$;
their explicit form is given below in (\ref{Krec}).
The scalar functions $V_{r}^{(k)}$ are basic separable potentials, see below for details.
\looseness=-1

The contravariant metric
tensors $G_{m}$ have the form \cite{mac2005}%
\[
G_{m}=L^{m}G_{0},\quad m\in\mathbb{Z},\quad G_{0}=\mathrm{diag}\left(  \frac{1}{\Delta_{1}%
},\ldots,\frac{1}{\Delta_{n}}\right),%
\]
where $\Delta_{i}=%
{\textstyle\prod\limits_{j\neq i}} (\lambda^{i}-\lambda^{j})$, and
$L=\mathrm{diag}(\lambda^{1},\ldots,\lambda ^{n})$ is a
$(1,1)$-tensor on $\mathcal{Q}$ called a {\em special conformal
Killing tensor} \cite{c}. Interestingly enough, these metrics also
appear in the theory of local Hamiltonian structures for systems
of hydrodynamic type, see e.g.\ \cite{fp, mokh} and references
therein. More general separable metrics can be obtained from $G_m$
via the so-called $k$-hole deformations \cite{mac2005, ms2006}.
For $m=0,\dots,n$ the (contravariant)
metrics $G_{m}$ are flat \cite{fp}, and in what follows we shall restrict
ourselves to considering these metrics only.

For the sake of comparison notice that for the metric $\bar G$ given by
\be\label{ell} \bar G=\left(\prod_{i=1}^n (L-\alpha_i)\right) G_0,
\ee
where $\alpha_i$ are nonzero constants such that $\alpha_i\neq
\alpha_j$ for $i\neq j$ and $\alpha_1 < \cdots <\alpha_n$, the
separation curve reads
\[
H_{1}\lambda^{n-1}+H_{2}\lambda^{n-2}
+\cdots+H_{n}=\frac{1}{2}{\textstyle\prod\limits_{j=1}^n}
(\lambda-\alpha_{j})%
\mu^{2}+\lambda^{k}.
\]
Then the $\lambda$'s turn out to be nothing
but the well-known (see e.g.\ \cite{kal} and references therein)
elliptic coordinates related to the flat coordinates $x^k$
by the formula
\be\label{elco}
(x^k)^2=4 \frac{{\prod\limits_{j=1}^n}
(\alpha_{k}-\lambda^{j})}{{\prod\limits_{j=1,j\neq k}^n}
(\alpha_{k}-\alpha_{j})},\qquad k=1,\dots,n,
\ee
where $\alpha_1< \lambda^1 < \alpha_2 < \lambda^2 < \cdots < \alpha_n <\lambda^n$.
In the $x$-coordinates $\bar G$ becomes the $n\times n$ unit matrix.
Moreover, all separable coordinate systems in $\mathbb{R}^n$ endowed with an Euclidean metric
can be obtained \cite{kal} as degenerations of (\ref{elco}). However,
unlike $\bar G$, the metrics $G_m$, $m=0,\dots,n$, are in general
pseudo-Euclidean (for more details, see the end of Section 1),
so this result does not apply to them.
\looseness=-2

The Killing tensors $K_{r}$ from (\ref{Ham}) are
diagonal in the $\lambda$-coordinates and have the following form
\cite{ben93}:
\looseness=-1
\begin{equation}
K_{1}=\mathbb{I},\quad
K_{r}=\sum_{k=0}^{r-1}(-1)^{k}\sigma_{k}(\lambda)L^{r-1-k},\quad
r=2,\dots,n,\label{Krec}%
\end{equation}
where $\mathbb{I}$ is the $n\times n$ unit matrix, and
$\sigma_{k}=\sigma_{k}(\lambda)$ are symmetric polynomials in the variables
$\lambda^{1},\dots
,\lambda^{n}$ ($\sigma_{0}=1$, $\sigma_{1}=\sum_{i=1}^{n}\lambda^{i}%
,\dots,\sigma_{n}=\lambda^{1}\lambda^{2}\cdots\lambda^{n}$).
They are related to coefficients of the
characteristic polynomial of the tensor $L$ as follows:
\begin{equation}
\det(\xi\mathbb{I}-L)=\sum\limits_{i=0}^{n}(-1)^{i}\sigma_{i}(\lambda)\xi
^{n-i}.\label{Viete}%
\end{equation}
The explicit form of geodesic Hamiltonians is \cite{1}
\begin{equation}
E_{m,r}\equiv H_r^{(m,0)}=\frac{1}{2}\mu^{T}K_{r}G_{m}\mu=\frac{(-1)^{r-1}}{2}\sum_{i=1}%
^{n}\frac{\partial\sigma_{r}}{\partial\lambda^{i}}\frac{(\lambda^{i})^{m}%
}{\Delta_{i}}\mu_{i}^{2}.\label{E}%
\end{equation}

The potentials $V_{r}^{(k)}$ in the Hamiltonians (\ref{Ham})
can be obtained from the following recursion relation~\cite{mac2005}:
\looseness=-2
\begin{equation}
V_{r}^{(k)}=V_{r+1}^{(k-1)}+V_{r}^{(n)}V_{1}^{(k-1)},\qquad V_{r}%
^{(n)}=-(-1)^{r}\sigma_{r}(\lambda),\qquad k\in\mathbb{Z},\label{rekup}
\end{equation}
with the initial condition%
\begin{equation}
V_{r}^{(0)}=\delta_{r,n}\qquad r=1,\ldots,n.\label{in}%
\end{equation}
Here we tacitly assume that $V_{r}^{(k)}\equiv0$ for$\ \ r<1\ \ $or$\
\ r>n\ $.

The recursion (\ref{rekup}) can be reversed. The inverse recursion is given by%
\begin{equation}
\fl V_{r}^{(k)}=V_{r-1}^{(k+1)}+V_{r}^{(-1)}V_{n}^{(k+1)},\qquad V_{r}%
^{(-1)}=(-1)^{n-r}\frac{\sigma_{r-1}(\lambda)}{\sigma_{n}(\lambda)},\qquad k\in
\mathbb{Z},\quad r=1,\ldots,n. \label{rekdown}%
\end{equation}
Hence, the first nonconstant potentials are $V_{r}^{(n)}$ for $k>0$ and
$V_{r}^{(-1)}$ for $k<0$, respectively. The reader may wish to compare
these potentials with their counterparts for the metric $\bar G$ (\ref{ell})
in the $x$-coordinates (\ref{elco}), see \cite{st}.
\looseness=-1

\section{Flat coordinates}
Our first step in the construction of natural coordinates
is to perform the canonical transformation from the $(\lambda,\mu)$-
to the $(q,p)$-coordinates defined as follows \cite{bl04}:
\begin{equation}
q^{i}=(-1)^{i}\sigma_{i}(\lambda),\quad p_{i}=-\sum\limits_{k=1}^{n}%
(\lambda^{k})^{n-i}\mu_{k}/\Delta_{k},\quad i=1,\dots,n.\label{first}%
\end{equation}
Notice that $q^{r}$ are nothing but coefficients of the
characteristic polynomial of $L$ (\ref{Viete}).
In the $(q,p)$-coordinates we have \cite{bl04}
\[
L_{j}^{i}=-\delta^1_j q^i+\delta^{i+1}_j,\quad
(G_{m})^{rs}=\left\{
\ba{l}
\sum\limits_{j=0}^{n-m-1}q^{j}\delta_{n-m+j+1}^{r+s},\quad r,s=1,\dots,n-m,\\[5mm]
-\hspace{-5mm}\sum\limits_{j=n-m+1}^{n}q^{j}\delta_{n-m+j+1}^{r+s},
\quad r,s=n-m+1,\dots,n,\\[5mm]
0\quad\mbox{otherwise.}
\ea
\right.
\]
where we set $q^0\equiv 1$ for convenience and $m=0,\dots, n$.

An important advantage of these new coordinates is polynomiality
of geodesic Hamiltonians in $p$'s and $q$'s \cite{bl04}:
\[
\ba{l}
E_{m,1}=
\ds\frac{1}{2}\sum\limits_{k=0}^{n-m-1}q^{k}\sum\limits_{j=k+1}%
^{n-m}p_{j}p_{n-m+k-j+1}
-\frac{1}{2}\sum\limits_{k=1}^{m}q^{n-m+k}%
\sum\limits_{j=1}^{k}p_{n-m+j}p_{n-m+k-j+1}.\label{E1}%
\ea
\]


At the second step, we fix the value of $m$
and perform a canonical transformation
from the $(q,p)$- to the $(r,s)$-coordinates defined by means
of the formulas
\begin{equation}\label{second}
\begin{array}{ll}
q^{i}  &
=r^{i}+\ds{\frac{1}{4}}\sum\limits_{j=1}^{i-1}r^{j}r^{i-j},\quad
i=1,\dots,n-m,
\\
q^{i}  &  =-\ds{\frac{1}{4}}\sum\limits_{j=i}^{n}r^{j}r^{n-j+i},\quad
i=n-m+1,\dots,n,\\
s_{k}  &  =\ds\sum\limits_{i=1}^{n}\displaystyle\frac{\partial q^{i}}{\partial r^{k}}p_{i}%
,\quad k=1,...,n.
\end{array}
\end{equation}
It is straightforward to verify that the following assertion holds:
\begin{theorem}\label{flath}
For any given $m$, $0\leq m\leq n$,
the metrics $G_{m}$ in the coordinates $r^i$ defined by (\ref{second})
takes the form
\begin{equation}
(G_{m})^{kl}=\left(
\delta_{n-m+1}^{k+l}+\delta_{2n-m+1}^{k+l}\right),
\end{equation}
and in the $(r,s)$-coordinates (\ref{second}) we have
\begin{equation}
E_{m,1}=\frac{1}{2}\left(\sum\limits_{j=1}^{n-m}s_{j}s_{n-m+1-j}%
+\sum\limits_{j=n-m+1}^{n}s_{j}s_{2n-m+1-j}\right)  . \label{E2}%
\end{equation}
The tensor $L$ in the coordinates $r^i$ takes the form:
\[
\ba{l}
\ds\fl\mbox{for $m<n$:}\quad L^i_j=\delta^{i+1}_j(1-\delta^i_{n-m}) -\frac12 r^i \delta^1_j
-\frac12 r^{n-j-m+1+n[(j+m-1)/n]} \delta^i_{n-m}\\[5mm]
\ds\fl\mbox{for $m=n$:}\quad L^i_j=\delta^{i+1}_j+\frac14 r^i r^{n-j+1}.
\ea
\]
\end{theorem}
Here $[k]$ denotes the greatest integer less than or equal to $k$
and $\delta_i^j$ is 
the Kronecker delta.
\looseness=-1

Notice that although the canonical coordinates $(r,s)$ are still
nonorthogonal, the metric tensor $G_{m}$ is constant in these coordinates.
In order to bring $G_m$ into canonical form, with $+1$ and $-1$ at the diagonal
and zeros off the diagonal, we should perform one more
%
canonical transformation from the $(r,s)$- to the
$(x,\pi)$-coordinates
defined as follows (here $d\equiv [(n-m)/2]$):
\begin{theorem}\label{orth}
For any given $m$, $0\leq m\leq n$,
the transformation defined by the formulas
\begin{equation}%
\begin{array}
[c]{l}%
\fl \pi_{k}=\sum\limits_{i=1}^{n}\displaystyle\frac{\partial
r^{i}}{\partial x^{k}}s_{i}
,\quad k=1,\dots,n,\\[5mm]
\fl r^{i}=(x^{i}+x^{n-i+1})/\sqrt{2},\quad i=1,\dots,d,\\
\fl r^{i}=(x^{n-m-i+1}-x^{m+i})/\sqrt{2},\quad
i=n-m-d+1,\dots,n-m,\\
\fl r^{i}=(x^{i-n+m+d}+x^{2n-m-d-i+1})/\sqrt{2},\quad i=n-m+1,\dots,n-m+[m/2],\\
\fl r^{i}=(x^{n+1+d-i}-x^{i-d})/\sqrt{2},\quad
i=n+1-[m/2],\dots,n,\\
\fl\mbox{if $n-m$ is odd and $m$ is even then
$r^{d+1}=x^{d+[m/2]+1}$,}\\
\fl\mbox{if $n-m$ is even and $m$ is odd then
$r^{n-m+[m/2]+1}=x^{d+[m/2]+1}$,}\\
\fl\mbox{if both $n-m$ and $m$ are odd then
$r^{d+1}=x^{d+[m/2]+1}$ and $r^{n-m+[m/2]+1}=x^{d+[m/2]+2}$,}
\end{array}
\label{third}%
\end{equation}
brings the metrics $G_{m}$ into the canonical form
\begin{equation}\label{canon}
G_{m}^{ij}=\left\{
\begin{array}
[c]{l}%
+1,\mbox{if $i=j$ and
$i=1,\dots,n-[(n-m)/2]-[m/2]$},\\
-1,\mbox{if $i=j$ and $i=n-[(n-m)/2]-[m/2]+1,\dots,n$},\\
0\quad\mbox{otherwise.}
\end{array}
\right.
\end{equation}
and we have
\begin{equation}
\fl E_{m,1}=\frac{1}{2}\left( \sum\limits_{j=1}^{n-d-[m/2]}\pi_{j}^{2}%
-\sum\limits_{j=n-d-[m/2]+1}^{n}\pi_{j}^{2}\right)  ,\quad
m=0,\dots,n.
\label{E3}%
\end{equation}
\end{theorem}
Thus, $x^i$ are orthogonal coordinates for the metric $G_m$, cf.\
(\ref{E3}), and hence $x^{j}$ and $\pi_j$ provide
\emph{natural coordinates} for all Hamiltonians $H_{1}^{(m,k)}=E_{m,1}%
+V_{1}^{(k)}$, $m=0,\dots,n$, $k\in\mathbb{Z}$, from the Benenti
class (\ref{SC1}). The relationship of the $q$- and
$x$-coordinates can be readily recovered from direct comparison of the
coefficients of characteristic polynomial for $L$ in the
corresponding coordinate frames.
Note that the Hamiltonians $H_{1}^{(m,k)}$ with $k=-m,...,2n-m-2$
for $m=0,...,n-1$ and $H_{1}^{(n,k)}$ with $k=-n+2,...,n$ for
$m=n,n+1$ are maximally superintegrable \cite{bl04}. 

As a final remark notice that, unlike $\bar G$ (\ref{ell}), which is Euclidean,
the metrics $G_m$, $m=0,\dots,n$, are in general pseudo-Euclidean
with the signature $(n-[(n-m)/2]-[m/2], [(n-m)/2]+[m/2])$, i.e., there are
$n-[(n-m)/2]-[m/2]$ positive and $[(n-m)/2]+[m/2]$ negative
entries in the canonical form (\ref{canon}) of $G_m$.

\section{Examples}
Let us illustrate our results for $n=4$. In the
$(q,p)$-coordinates we have
\[
G_{0}=\left(
\begin{array}
[c]{cccc}%
0 & 0 & 0 & 1\\
0 & 0 & 1 & q^{1}\\
0 & 1 & q^{1} & q^{2}\\
1 & q^{1} & q^{2} & q^{3}%
\end{array}
\right)  ,\ G_{1}=\left(
\begin{array}
[c]{cccc}%
0 & 0 & 1 & 0\\
0 & 1 & q^{1} & 0\\
1 & q^{1} & q^{2} & 0\\
0 & 0 & 0 & -q^{4}%
\end{array}
\right)  ,\ G_{2}=\left(
\begin{array}
[c]{cccc}%
0 & 1 & 0 & 0\\
1 & q^{1} & 0 & 0\\
0 & 0 & -q^{3} & -q^{4}\\
0 & 0 & -q^{4} & 0
\end{array}
\right),
\]%
\[
G_{3}=\left(
\begin{array}
[c]{cccc}%
1 & 0 & 0 & 0\\
0 & -q^{2} & -q^{3} & -q^{4}\\
0 & -q^{3} & -q^{4} & 0\\
0 & -q^{4} & 0 & 0
\end{array}
\right)  ,\ G_{4}=\left(
\begin{array}
[c]{cccc}%
-q^{1} & -q^{2} & -q^{3} & -q^{4}\\
-q^{2} & -q^{3} & -q^{4} & 0\\
-q^{3} & -q^{4} & 0 & 0\\
-q^{4} & 0 & 0 & 0
\end{array}
\right), L=\left(
\begin{array}
[c]{cccc}%
-q^{1} & 1 & 0 & 0\\
-q^{2} & 0 & 1 & 0\\
-q^{3} & 0 & 0 & 1\\
-q^{4} & 0 & 0 & 0
\end{array}
\right),
\]
with the simplest nontrivial potentials being
\[
\fl
V_{1}^{(-3)}=(q^{2}q^{4}-(q^{3})^{2})/(q^{4})^{3},\quad V_{1}^{(-2)}%
=q^{3}/(q^{4})^{2},\quad V_{1}^{(-1)}=1/q^{4},\quad
V_{1}^{(4)}=-q^{1},
\]%
\[
\fl
V_{1}^{(5)}=-q^{2}+(q^{1})^{2},\quad V_{1}^{(6)}=-q^{3}+2q^{1}q^{2}-(q^{1}%
)^{3},\quad V_{1}^{(7)}=-q^{4}+2q^{1}q^{3}+(q^{2})^{2}-3(q^{1})^{2}q^{2}%
+(q^{1})^{4}.
\]
For $m=0$ in the $(r,s)$-coordinates we have
\[
q^{1}=r^{1},\ \ \ q^{2}=\frac{1}{4}(r^{1})^{2}+r^{2},\ \ \ q^{3}=\frac{1}%
{2}r^{1}r^{2}+r^{3},\ \ \ q^{4}=\frac{1}{2}r^{1}r^{3}+\frac{1}{4}(r^{2}%
)^{2}+r^{4},
\]%
\[
G_{0}=\left(
\begin{array}
[c]{llll}%
0 & 0 & 0 & 1\\
0 & 0 & 1 & 0\\
0 & 1 & 0 & 0\\
1 & 0 & 0 & 0
\end{array}
\right)  ,\ \ L=\left(
\begin{array}
[c]{cccc}%
-\frac{1}{2}r^{1} & 1 & 0 & 0\\
-\frac{1}{2}r^{2} & 0 & 1 & 0\\
-\frac{1}{2}r^{3} & 0 & 0 & 1\\
-r^{4} & -\frac{1}{2}r^{3} & -\frac{1}{2}r^{2} & -\frac{1}{2}r^{1}%
\end{array}
\right),
\]
while in the $(x,\pi)$-coordinates
\[
\ba{l} \fl q^{1}=\ds\frac{1}{\sqrt{2}}\left(x^{1}+x^4\right),\quad
\fl q^{2}=\ds\frac18\left(x^1+x^4\right)^2+\ds\frac{1}{\sqrt{2}}\left(x^2+x^3\right),\\[4mm]
\fl q^{3}=\ds\frac14\left(x^1 x^2+x^1 x^3+x^4 x^2+x^4 x^3\right)+\ds\frac{1}{\sqrt{2}}\left(x^2-x^3\right),\\[4mm]
\fl q^{4}=\ds\frac14\left(x^1 x^2-x^1 x^3+x^4 x^2-x^4
x^3\right)+\ds\frac18\left(x^2+x^3\right)^2
+\ds\frac{1}{\sqrt{2}}\left(x^1-x^4\right),
\ea
\]%
\[
\fl G_{0}=\left(
\begin{array}
[c]{cccc}%
1 & 0 & 0 & 0\\
0 & 1 & 0 & 0\\
0 & 0 & -1 & 0\\
0 & 0 & 0 & -1
\end{array}
\right),\ L=\frac{1}{\sqrt{2}}\left(
\begin{array}
[c]{cccc}%
-x^{1} & -\frac{1}{2}x^{2}+\frac{1}{\sqrt{2}} &
\frac{1}{2}x^{3}+\frac
{1}{\sqrt{2}} & \frac{1}{2}x^{4}-\frac{1}{2}x^{1}\\
-\frac{1}{2}x^{2}+\frac{1}{\sqrt{2}} & \frac{1}{\sqrt{2}} &
-\frac{1}{\sqrt
{2}} & -\frac{1}{2}x^{2}-\frac{1}{\sqrt{2}}\\
-\frac{1}{2}x^{3}-\frac{1}{\sqrt{2}} & \frac{1}{\sqrt{2}} &
-\frac{1}{\sqrt
{2}} & -\frac{1}{2}x^{3}+\frac{1}{\sqrt{2}}\\
\frac{1}{2}x^{1}-\frac{1}{2}x^{4} &
\frac{1}{2}x^{2}+\frac{1}{\sqrt{2}} &
-\frac{1}{2}x^{3}+\frac{1}{\sqrt{2}} & -x^{4}%
\end{array}
\right),
\]
and, for instance, the Hamiltonian $H_1^{(0,6)}$ reads
\[
H_1^{(0,6)}=\ds\frac12\left(\pi_1^2+\pi_2^2-\pi_3^2-\pi_4^2\right)
+\ds\frac34\left(x^1 x^2+x^1 x^3+x^4 x^2+x^4 x^3\right)
+\ds\frac{1}{\sqrt{2}}\left(x^3-x^2\right)
-\ds\frac{1}{4\sqrt{2}}\left(x^1+x^4\right)^3.
\]
For another choice $m=n-1=3$ in the $(r,s)$-coordinates we obtain
\[
q^{1}=r^{1},\quad q^{2}=-\frac{1}{4}(r^{3})^{2}-\frac{1}{2}r^{2}%
r^{4},\quad q^{3}=-\frac{1}{2}r^{3}r^{4},\quad
q^{4}=-\frac{1}{4}(r^{4})^{2},
\]%
\[
G_{3}=\left(
\begin{array}
[c]{llll}%
1 & 0 & 0 & 0\\
0 & 0 & 0 & 1\\
0 & 0 & 1 & 0\\
0 & 1 & 0 & 0
\end{array}
\right)  ,\ \ L=\left(
\begin{array}
[c]{cccc}%
-r^{1} & -\frac{1}{2}r^{4} & -\frac{1}{2}r^{3} & -\frac{1}{2}r^{2}\\
-\frac{1}{2}r^{2} & 0 & 1 & 0\\
-\frac{1}{2}r^{3} & 0 & 0 & 1\\
-\frac{1}{2}r^{4} & 0 & 0 & 0
\end{array}
\right).
\]
Hence in the $(x,\pi)$-coordinates we have
\[
\ba{l} \fl q^{1}=x^{2},\quad \fl
q^{2}=-\ds\frac{1}{4}\left((x^1)^2+(x^3)^2-(x^4)^2\right),\quad
\fl q^{3}=-\ds\frac{1}{2\sqrt{2}}\ x^3\left(x^1-x^4\right),\quad
\fl q^{4}=-\ds\frac{1}{8}\left(x^1-x^4\right)^2,\quad \ea
\]
\[
G_{3}=\left(
\begin{array}
[c]{cccc}%
1 & 0 & 0 & 0\\
0 & 1 & 0 & 0\\
0 & 0 & 1 & 0\\
0 & 0 & 0 & -1
\end{array}\right),\quad L=\left(
\begin{array}
[c]{cccc}%
0 & -\frac{1}{2}x^{1} & \frac{1}{\sqrt{2}} & 0\\
-\frac{1}{2}x^{1} & -x^{2} & -\frac{1}{2}x^{3} & \frac{1}{2}x^{4}\\
\frac{1}{\sqrt{2}} & -\frac{1}{2}x^{3} & 0 & -\frac{1}{\sqrt{2}}\\
0 & -\frac{1}{2}x^{4} & \frac{1}{\sqrt{2}} & 0
\end{array}
\right),
\]
and, for example, the Hamiltonian $H_1^{(3,6)}$ 
takes the form
\[
H_1^{(3,6)}=\ds\frac12\left(\pi_1^2+\pi_2^2+\pi_3^2-\pi_4^2\right)
-\ds\frac12
\left((x^1)^2+2(x^2)^2+(x^3)^2-(x^4)^2\right)x^2+\ds\frac{1}{2\sqrt{2}}\left(x^1-x^4\right)
x^3.
\]
Let us mention that for $m=0$ and arbitrary dimension $n$
the potentials $V_1^{(2n+1)}$ correspond to stationary flows of the
KdV soliton hierarchy while for $m=n-1$ the potentials $V_1^{(-n)}$
are related to stationary flows of the Harry Dym hierarchy
\cite{mb, blasak1, blasak}.
Further applications of the flat coordinates
presented in Theorems \ref{flath} and \ref{orth} can be found in \cite{se06}.

\section*{Acknowledgments}
This research was partially supported by the Czech Grant Agency
(GA\v{C}R) under grant No.~201/04/0538, by the Ministry of Education,
Youth and Sports of the Czech Republic (M\v SMT \v CR) under grant MSM 4781305904,
and by the Polish State Committee For Scientific Research (KBN)
under the KBN Research Grant No.\ 1 PO3B 111 27. M.B. is pleased to acknowledge
kind hospitality of the Mathematical Institute of Silesian
University in Opava.
\looseness=-1


\end{document}